\documentclass[a4paper,reqno]{amsart}
\usepackage{amssymb,latexsym}

\begin{document}

\title[Duality-Symmetric Gravity and Supergravity]
{Duality-Symmetric Gravity and Supergravity:\\
testing the PST approach}

\author{Alexei~J. Nurmagambetov}

\address{A.I. Akhiezer Institute for Theoretical Physics\\
NSC ``Kharkov Institute of Physics and Technology"\\
Kharkov, 61108, Ukraine}

\email{ajn@kipt.kharkov.ua}
%\thanks{Work supported in part ...}
%\subjclass{04.65.+e, 04.50.+h, 11.25.Mj}
\date{\today}

\begin{abstract}
Drawing an analogy between gravity dynamical equation of motion
and that of Maxwell electrodynamics with an electric source we
outline a way of appearance of a dual to graviton field. We
propose a dimensional reduction ansatz for the field strength of
this field which reproduces the correct duality relations between
fields arising in the dimensional reduction of D-dimensional
gravity action to D-1 dimensions. Modifying the PST approach we
construct a new term entering the action of D=11 duality-symmetric
gravity and by use of the proposed ansatz we confirm the relevance
of such a term to reproduce the correct duality-symmetric
structure of the reduced theory. We end up extending the results
to the bosonic sector of D=11 supergravity.
\end{abstract}

\maketitle

\tableofcontents

\section{Introduction}

Recent progress in studying the hidden symmetry group of
Superstring/M-theory \cite{westjhep00}--\cite{nurE8} reveals
necessity of taking seriously into account a ``graviton dual"
field. Dynamics of such a field has been intensively studied in
literature in the linearized limit \cite{hull}, \cite{westcqg01},
\cite{westcqg03}, \cite{boulanger}, \cite{cmu} together with
studying the actions that lead in the same approximation to the
equation of motion for the ``graviton dual" field. The attempts to
extend the linear theory to a non-linear one faced the troubles
that were summarized in the ``no-go" theorems \cite{boulanger}
which forbid the action for interacting theory solely in terms of
the ``graviton dual" field. This situation is very similar to that
of constructing eleven-dimensional supergravity with a six-index
photon field \cite{ntn}, \cite{df} where the simple arguments
\cite{cjlp0} lead to the conclusion on impossibility of
constructing the gauge sector of theory solely in terms of the
six-index photon. However a way to go beyond these arguments is to
consider a duality-symmetric theory of D=11 supergravity with an
original three-index photon field and with the ``dual-to-three
index photon" field that enter the action on equal footing
\cite{bbs} (see also \cite{alwis} for an early attempt). Here we
will follow the same way and will closely inspect a possibility to
construct the non-linear theory of gravity and supergravity with
the ``graviton dual" field standing on a point of the
duality-symmetric formulation. The arguments in favor of such a
consideration are as follows. First, we recall the close
connection between D=11 supergravity and type IIA theory that
follows from the D=11 supergravity by dimensional reduction.
Importance of having the ``graviton dual" field in D=11
supergravity for D=10 type IIA theory has been pointed out as in
the purely algebraic aspect \cite{westcqg01} as well as in the
context of searching for the complete duality-symmetric version of
D=11 supergravity \cite{nur} involving the ``graviton dual" field.
As well as recovering the correct algebra for a non-linear
realization of type IIA supergravity requires of having a
generator corresponding to the graviton dual partner in its
eleven-dimensional counterpart \cite{westcqg03} this dual field is
required for recovering in a straightforward way the complete
duality-symmetric formulation of type IIA supergravity \cite{bns}.
Second, since the duality-symmetric version of type IIA
supergravity was realized on the ground of Pasti-Sorokin-Tonin
technique \cite{pst}, it is naturally to expect the PST-like
structure of additional terms in D=11 supergravity action which
encode the duality relations between the graviton and its dual
partner. As we have known the structure of the duality-symmetric
type IIA supergravity \cite{bns} we can therefore use of this
advantage for deducing an ansatz for the dimensional reduction of
the ``graviton dual" field. Finally, we can try to apply the
ansatz for constructing the additional terms in D=11 supergravity
action which after reduction will reproduce the correct
duality-symmetric structure of D=10 type IIA supergravity. To this
end we shall take into account the observations that have been
previously made under construction of duality-symmetric theories
for different sub-sectors of maximal higher-dimensional
supergravities \cite{bbs}, \cite{dls}, \cite{dlt}, \cite{bns}.

It has to be emphasized that as a first step towards completing
our task we shall find, without an appeal to a method of
constructing the action, a convenient representation of the
gravity equation of motion in a way that allows us to present the
latter as the Bianchi identity for a dual field. It turns out to
be convenient for this purpose to write down the gravity equation
of motion in a form which is similar to the dynamics of Maxwell
theory with a source. As soon as such a representation is
recovered it suggests a way of extracting the dual field and to
apply the PST formalism to construct the action from which the
duality relations between graviton and its dual partner will
follow as equations of motion.

For the sake of simplicity we are getting started our quest of the
complete duality-symmetric formulation of D=11 supergravity with
pure gravity case extended hereafter the results to the bosonic
sector of D=11 supergravity. The scheme we propose does not depend
on the space-time dimension $D\ge 4$ where gravity has dynamical
degrees of freedom, and since the gravity is the key ingredient of
supergravity the results can be applied for all theories of
supergravity in diverse dimensions though as we have mentioned
above we will end up with the dimension of space-time D=11 having
in mind mostly the applying to this case.

\section{Kaluza-Klein ansatz for graviton dual field}

Let us begin with introducing the graviton field described by a
vielbein $\hat{e}^{{\hat a}[1]}$ whose index $\hat{a}$ runs from
zero to $D-1$ and takes the value in the tangent space Lorentz
group $SO(1,D-1)$, and with introducing the dual to the vielbein
field $\hat{A}^{{\hat a}[D-3]}$. The way in which the latter
appears is as follows. The first order action for the pure gravity
is
\begin{equation}\label{EH}
S_{EH}=\int_{{\mathcal M}^{D}}\, \hat{R}^{\hat{a}\hat{b}}\wedge
\hat{\Sigma}_{\hat{a}\hat{b}},
\end{equation}
where
$\hat{R}^{\hat{a}\hat{b}}=d\hat{\omega}^{\hat{a}\hat{b}}-{\hat{\omega}}^{\hat
a}_{~\hat{c}}\wedge \hat{\omega}^{\hat{c}\hat{b}}$ is the
curvature 2-form and
\begin{equation}\label{Sig}
\hat{\Sigma}_{{\hat a}_1\dots {\hat a}_n}={1\over
(D-n)!}\epsilon_{{\hat a}_1\dots {\hat a}_{D}}{\hat e}^{{\hat
a}_{n+1}}\wedge\dots\wedge {\hat e}^{{\hat a}_{D}}
\end{equation}
is the $(D-n)$-form constructed out of vielbeins ${\hat e}^a$. The
equations of motion following from the action are
\begin{equation}\label{Rem}
\hat{\Sigma}_{{\hat a}{\hat b}{\hat c}}\wedge \hat{R}^{{\hat
b}{\hat c}}=0,
\end{equation}
\begin{equation}\label{Tem}
\hat{\Sigma}_{{\hat a}{\hat b}{\hat c}}\wedge \hat{T}^{\hat c}=0,
\end{equation}
where we have introduced the torsion 2-form $\hat{T}^{\hat
a}=d\hat{e}^{\hat a}-\hat{e}^{\hat b}\wedge{{\hat \omega}_{\hat
b}}^{~{\hat a}}$. The latter equation set the torsion to zero that
is the algebraic relation expressing the connection
$\hat{\omega}^{{\hat a}{\hat b}}$ through the vielbeins and their
derivatives. In the sequel we will be mostly concentrated on the
dynamical equation of motion.

It is easy to check that the Einstein equation \eqref{Rem} admits
the following representation
\begin{equation}\label{Rem1}
d({\hat \omega}^{{\hat b}{\hat c}}\wedge \hat{\Sigma}_{{\hat
a}{\hat b}{\hat c}})=\hat{\omega}^{{\hat b}{\hat c}}\wedge
d\hat{\Sigma}_{{\hat a}{\hat b}{\hat c}}+(-)^{D-3}~{{\hat
\omega}}^{\hat b}_{~{\hat d}}\wedge \hat{\omega}^{{\hat d}{\hat
c}}\wedge \hat{\Sigma}_{{\hat a}{\hat b}{\hat c}}.
\end{equation}
One can rewrite this equation in the form
\begin{equation}\label{fem}
d\hat{\ast} \hat{f}^{[2]}_{\hat a}=\hat{\ast} \hat{J}^{[1]}_{\hat
a}
\end{equation}
with $\hat{\ast} \hat{f}^{[2]}_{\hat a}\equiv {\hat \omega}^{{\hat
b}{\hat c}}\wedge \hat{\Sigma}_{{\hat a}{\hat b}{\hat c}}$, that
looks very much like the equation of motion for Maxwell
electrodynamics with an electric source. This analogy has been
pointing out in literature for a long time (cf. e.g.
\cite{thirring}, \cite{blau}). However, the difference between the
electrodynamics equation of motion and \eqref{fem} is apparent.
The former contains a ``bare" potential $A^{[1]}$ whereas the
latter does not, since the r.h.s. and the l.h.s. of \eqref{fem}
depend on the vielbeins and connection. The only vielbein is the
true dynamical field, hence one should resolve the connection
through the vielbeins using the torsion free constraint
\eqref{Tem}. The answer is
\begin{equation}\label{om}
{{\hat \omega}_{{\hat m}}}^{~~{\hat a}{\hat b}}={1\over
2}\hat{e}^{{\hat n}{\hat a}}\partial_{[{\hat m}}\hat{e}_{{\hat
n}]}^{\hat b}- {1\over 2}\hat{e}^{{\hat n}{\hat
b}}\partial_{[{\hat m}}\hat{e}_{{\hat n}]}^{\hat a}-{1\over
2}\hat{e}_{{\hat m}{\hat c}}\hat{e}^{{\hat n}{\hat
a}}\hat{e}^{{\hat s}{\hat b}}
\partial_{[{\hat n}}\hat{e}_{{\hat s}]}^{\hat c},
\end{equation}
where the indices from the second middle of Latin alphabet are the
curved ones. Substituting this expression into the l.h.s. of
\eqref{Rem1} we arrive at
\begin{equation}\label{Rem2}
d(\hat{\omega}^{{\hat b}{\hat c}}\wedge\hat{\Sigma}_{{\hat a}{\hat
b}{\hat c}})=\frac{1}{{\boldsymbol \alpha}_D} d(\hat{\ast}
d\hat{e}_{\hat a})+d\hat{S}^{[D-2]}_{\hat a},
\end{equation}
where we have introduced the form $\hat{S}^{[D-2]}_{\hat a}$ which
is a function of vielbeins and their derivatives and the numerical
coefficient ${{\boldsymbol \alpha}_D}$ which takes $\pm 1$ in
dependence on the space-time signature setting and the Hodge star
definition. It is worth mentioning that the r.h.s. of eq.
\eqref{fem} is nothing but the Landau-Lifshitz pseudo-tensor which
is a conserved ``current" and therefore admits the following
representation
\begin{equation}\label{J}
\hat{\omega}^{{\hat b}{\hat c}}\wedge d\hat{\Sigma}_{{\hat a}{\hat
b}{\hat c}}+(-)^{D-3}~{{\hat \omega}}^{\hat b}_{~{\hat d}}\wedge
\hat{\omega}^{{\hat d}{\hat c}}\wedge \hat{\Sigma}_{{\hat a}{\hat
b}{\hat c}}:= \hat{\ast}\hat{J}^{[1]}_{\hat a}=d\hat{\ast}
\hat{G}^{[2]}_{\hat a}.
\end{equation}
Taking into account \eqref{Rem2} and \eqref{J}, eq. \eqref{Rem1}
can be rewritten as
\begin{equation}\label{Rem3}
d(\hat{\ast} d\hat{e}_{\hat
a})=\hat{\ast}\hat{\tilde{J}}^{[1]}_{\hat a},
\end{equation}
with
\begin{equation}\label{tJ}
\hat{\ast}\hat{\tilde{J}}^{[1]}_{\hat
a}=d\hat{\ast}\hat{\tilde{G}}^{[2]}_{\hat a},
\end{equation}
where we have denoted $\hat{\tilde{G}}^{[2]}_{\hat
a}=\boldsymbol{\alpha}_D(\hat{G}^{[2]}_{\hat
a}-\hat{\ast}\hat{S}^{[D-2]}_{\hat a})$. Notice that
$\hat{\tilde{G}}^{[2]}_{\hat a}$ is also a function of vielbeins
and their derivatives.

One more comment is needed before proceeding further. The Einstein
equation \eqref{Rem} is generally covariant while its
representations \eqref{Rem1}, \eqref{fem}, \eqref{Rem3} are
recorded w.r.t. the usual (non-covariant) derivatives that could
lead to the conclusion on their non-covariance. This puzzle can be
resolved through the observation (cf. for instance \cite{schr})
that the r.h.s. of eqs. \eqref{Rem1}, \eqref{fem}, \eqref{Rem3} is
not a true tensor, but a pseudo-tensor, hence the non-covariance
of the l.h.s. of the above mentioned eqs. is compensated by the
pseudo-tensor character of the r.h.s. leaving nevertheless eqs.
\eqref{Rem1}, \eqref{fem}, \eqref{Rem3} to be covariant. Another
way to see that is, for instance, to notice the relation
$d(\hat{\ast}d\hat{e}_{\hat a}-\hat{\ast}\hat{\tilde
G}^{[2]}_{\hat a})=\boldsymbol{\alpha}_D
(-)^{D-3}\Sigma_{\hat{a}\hat{b}\hat{c}}\wedge
\hat{R}^{\hat{b}\hat{c}}$.

Now we have the almost complete analogy between gravity and
Maxwell theory with an electric source in the sense of
representing the former through the only true potentials. The dual
to the $d\hat{e}_{\hat a}$ field strength is
\begin{equation}\label{Fam}
\hat{F}^{[D-2]}_{\hat a}=d\hat{A}^{[D-3]}_{\hat
a}+\hat{\ast}\hat{\tilde{G}}^{[2]}_{\hat a}\equiv \hat{\mathbb
F}^{[D-2]}_{\hat a}+\hat{\ast}\hat{\tilde{G}}^{[2]}_{\hat a},
\end{equation}
and the ``graviton dual" field has appeared.

Since a part of the dual to the graviton field will become dual
fields to a dilaton and a Kaluza-Klein vector field after
dimensional reduction from $D$ to $D-1$, we will use this fact to
deduce the reduction ansatz for the field strength of
$\hat{A}^{{\hat a}[D-3]}$ that will reproduce the correct
structure of the duality-symmetric gravity in $D-1$ space-time
dimensions. After establishing the ansatz we will try to apply it
for lifting the known $D-1$-dimensional action up to the
$D$-dimensional space-time.

Splitting the tangent space index as $\hat{a}=(a,z)$,
$a=0,\dots,D-2$ we choose the standard Kaluza-Klein ansatz for the
vielbein
\begin{equation}\label{KKe}
\hat{e}^{a[1]}=e^{\alpha \phi} e^{a[1]},\qquad
\hat{e}^{z[1]}=e^{\beta\phi}(dz+A^{[1]}),
\end{equation}
where $\phi$ denotes the dilaton field, $A^{[1]}$ stands for the
Kaluza-Klein vector field and $z$ is the direction of the
reduction. All the fields on the r.h.s. of eq. \eqref{KKe} are
independent on the reduction coordinate. The parameters $\alpha$
and $\beta$ are related to fixing the space-time dimensions $D$
and to each other via \cite{lpss}
\begin{equation}\label{ab}
\alpha^2={1\over 2(D-2)(D-3)},\qquad \beta=-(D-3)\alpha,
\end{equation}
that corresponds to the Einstein frame after reduction.

The vielbeins' dimensional reduction ansatz falls into a general
class of ans\"atze which determine the reduction of a
(Lorentz-valued) $n$-form
\begin{equation}\label{drn}
\hat{\Omega}^{(a,z)[n]}=\Omega^{(a,z)[n]}+\omega^{(a,z)[n-1]}\wedge
(dz+A^{[1]}).
\end{equation}
To specify the ansatz one has to determine the forms
$\Omega^{(a,z)[n]}$, $\omega^{(a,z)[n-1]}$ for each case under
consideration. The other relation which will be under the usage in
what follows is
\begin{align}\label{drhn}
\hat{\ast}~\hat{\Omega}^{(a,z)[n]}&=a_{(n)}e^{-2(n-1)\alpha\phi}\ast
\Omega^{(a,z)[n]}\wedge
(dz+A^{[1]})\notag\\&+a_{(n-1)}e^{2(D-n-1)\alpha\phi}\ast
\omega^{(a,z)[n-1]}.
\end{align}
The numerical coefficients $a_{(n)}$, $a_{(n-1)}$ taking the
values $\pm 1$ encode the information on the space-time signature
choice and the definition of the Hodge operator. To distinguish
$D$ and $(D-1)$-dimensional Hodge stars we have equipped the
former with a hat.

Let us now turn to the analysis of gravity equation of motion
\eqref{Rem3}. The equation reads, at least for trivial topology
setting,
\begin{equation}\label{edr11}
\hat{\ast}d \hat{e}^{{\hat a}[1]}=d\hat{A}^{{\hat
a}[D-3]}+\hat{\ast}\hat{\tilde G}^{{\hat a}[2]}=\hat{\mathbb
F}^{{\hat a}[D-2]}+\hat{\ast}\hat{\tilde G}^{{\hat a}[2]} .
\end{equation}
Splitting the tangent space index $\hat{a}$ one arrives at the
following relations
\begin{equation}\label{edr11a}
\hat{\ast}d \hat{e}^{a[1]}=d\hat{A}^{
a[D-3]}+\hat{\ast}\hat{\tilde G}^{a[2]}=\hat{\mathbb F}^{a [D-2]}+
\hat{\ast}\hat{\tilde G}^{a[2]},
\end{equation}
\begin{equation}\label{edr11z}
\hat{\ast}d \hat{e}^{z[1]}=d\hat{A}^{z[D-3]}+\hat{\ast}\hat{\tilde
G}^{z[2]}= \hat{\mathbb F}^{z [D-2]}+\hat{\ast}\hat{\tilde
G}^{z[2]}.
\end{equation}
Since the duality relations for the dilaton and the Kaluza-Klein
vector field is encoded into the latter equation let us begin our
analysis with \eqref{edr11z}.

Taking into account \eqref{KKe}, \eqref{drn}, \eqref{drhn} one
arrives at the following relation
\begin{align}\label{0edr10}
& a_{(2)}e^{(\beta-2\alpha)\phi}\ast dA^{[1]}\wedge
(dz+A^{[1]})-a_{(1)}e^{2(D-3)\alpha\phi+\beta\phi}\beta\ast
d\phi=\notag\\
&={\mathbb F}^{z[D-2]}+{\mathbb F}^{z[D-3]}\wedge
(dz+A^{[1]})\notag\\
&+a_{(2)}e^{-2\alpha\phi}\ast G^{z[2]}\wedge
(dz+A^{[1]})+a_{(1)}e^{2(D-3)\alpha\phi}\ast g^{z[1]},
\end{align}
which contains two independent parts
\begin{equation}\label{1edr10}
a_{(2)}e^{(\beta-2\alpha)\phi}\ast dA^{[1]}={\mathbb F}^{z[D-3]}+
a_{(2)}e^{-2\alpha\phi}\ast G^{z[2]},
\end{equation}
and
\begin{equation}\label{2edr10}
-a_{(1)}e^{2(D-3)\alpha\phi+\beta\phi}\beta\ast d\phi={\mathbb
F}^{z[D-2]}+ a_{(1)}e^{2(D-3)\alpha\phi}\ast g^{z[1]}.
\end{equation}
Here we have denoted $\hat{\mathbb F}^{z[D-2]}\equiv
d\hat{A}^{z[D-3]}={\mathbb F}^{z[D-2]}+{\mathbb F}^{z[D-3]}\wedge
(dz+A^{[1]})$.

Since by construction (cf. \eqref{Rem2}, \eqref{Rem3}, \eqref{tJ})
$\hat{\tilde G}^{z[2]}$ depends on veilbeins and their derivatives
$G^{z[2]}$ and $g^{z[1]}$ are completely determined by the
reduction ansatz \eqref{KKe}. Hence, our aim is to deduce the
reduction ansatz for $\hat{\mathbb F}^{z[D-2]}$, i.e. to determine
${\mathbb F}^{z[D-2]}$ and ${\mathbb F}^{z[D-3]}$, which will lead
to the correct duality relations between the fields.

The ansatz we propose for ${\mathbb F}^{z[D-3]}$ has the following
form
\begin{equation}\label{F8ans}
{\mathbb F}^{z[D-3]}=-a_{(2)} e^{-\beta
\phi}dA^{[D-4]}-a_{(2)}e^{-2\alpha\phi}\ast G^{z[2]}.
\end{equation}
It is easy to see that the substitution of this ansatz into
\eqref{1edr10} leads to the correct duality relation between the
Kaluza-Klein vector field and its dual partner
\begin{equation}\label{KKdr10}
e^{2(\beta-\alpha)\phi}\ast dA^{[1]}=-dA^{[D-4]},
\end{equation}
which can be extracted from the Kaluza-Klein vector field equation
of motion after the dimensional reduction of pure gravity from $D$
to $D-1$.

The other part forming the ansatz for $\hat{\mathbb F}^{z[D-2]}$
is
\begin{align}\label{F9ans}
{\mathbb F}^{z [D-2]}&=a_{(1)}e^{-\beta\phi}(dA^{[D-3]}-{3\over 4}
dA^{[D-4]}\wedge A^{[1]})+a_{(1)}e^{2(D-3)\alpha\phi+\beta\phi}
(1-\beta)\ast d\phi\notag\\
&-a_{(1)}e^{2(D-3)\alpha\phi}\ast g^{z[1]}.
\end{align}
Substituting the latter into \eqref{2edr10} one can find the
correct duality relation between the dilaton and its dual partner
\begin{equation}\label{ddr10}
\ast d\phi=-dA^{[D-3]}+{3\over 4} dA^{[D-4]}\wedge A^{[1]}.
\end{equation}

Dealing with eq. \eqref{edr11a} is quite a bit complicated. By use
of \eqref{KKe}, \eqref{drn}, \eqref{drhn} one obtains
$$
a_{(2)} e^{-2\alpha\phi}\ast d(e^{\alpha\phi}e^a)\wedge
(dz+A^{[1]})={\mathbb F}^{a [D-2]}+{\mathbb F}^{a [D-3]}\wedge
(dz+A^{[1]})
$$
\begin{equation}\label{edr10a}
+a_{(2)}e^{-2\alpha\phi}\ast G^{a [2]}\wedge
(dz+A^{[1]})+a_{(1)}e^{2(D-3)\alpha\phi}\ast g^{a [1]},
\end{equation}
from which it immediately follows
\begin{equation}\label{Fa9ans}
{\mathbb F}^{a [D-2]}=-a_{(1)} e^{2(D-3)\alpha\phi}\ast g^{a [1]}.
\end{equation}
The other equation that is contained     into \eqref{edr10a} is
\begin{equation}\label{e10a1}
a_{(2)} e^{-\alpha\phi}\ast(de^a-\alpha~ d\phi\wedge e^a)={\mathbb
F}^{a [D-3]}+a_{(2)}e^{-2\alpha\phi}\ast G^{a [2]}.
\end{equation}

To find the correct ansatz for ${\mathbb F}^{a [D-3]}$ let us
introduce a new form ${\mathfrak G}^{a [2]}$. This form is defined
by
\begin{equation}\label{bfGd}
d\ast{\mathfrak G}^{a [2]}=\ast {\mathfrak J}^{a [1]}
\end{equation}
with
\begin{equation}\label{bfJd}
\ast{\mathfrak J}^{[1]}_a=\omega^{bc}\wedge
d\Sigma_{abc}+(-)^{D-4}~{\omega^b}_{~d}\wedge \omega^{dc}\wedge
\Sigma_{abc}-d S^{[D-3]}_a+M^{[D-2]}_a.
\end{equation}
Here $\omega^{ab}$ is the $(D-1)$-dimensional connection one-form,
$\Sigma_{abc}$ stands for the lower-dimensional analog of
\eqref{Sig}, $S^{[D-3]}_a$ is the $(D-1)$-dimensional counterpart
of the form $\hat{S}^{[D-2]}_{\hat a}$ introduced in \eqref{Rem2}
that appears after resolving the lower-dimensional torsion free
constraint, and $M^{[D-2]}_a$ is the energy-momentum tensor of the
dilaton and of the Kaluza-Klein vector field which is a closed
form and therefore
\begin{equation}\label{Md}
M^{[D-2]}_a=d k^{[D-3]}_a.
\end{equation}
Taking the ansatz to be
\begin{equation}\label{Fa8ans}
{\mathbb F}^{a [D-3]}=-a_{(2)}e^{-\alpha\phi}(dA^{a
[D-4]}+\ast{\mathfrak G}^{a [2]}+\alpha \ast (d\phi\wedge
e^a)+e^{-\alpha\phi}\ast G^{a [2]})
\end{equation}
one can obtain the duality relation (cf. \eqref{edr11})
\begin{equation}\label{edr10}
\ast de^a=-(dA^{a [D-4]}+\ast{\mathfrak G}^{a [2]}),
\end{equation}
that reproduces the correct equation of motion for the graviton
field
\begin{equation}\label{grem10}
d(\ast de^a)=-\ast{\mathfrak J}^{a [1]}.
\end{equation}

To summarize, the ansatz we proposed so far is
\begin{align}\label{Fa9s}
&\hat{\mathbb F}^{a [D-2]}=-a_{(1)} e^{2(D-3)\alpha\phi}\ast g^{a
[1]}\notag \\
&-a_{(2)}e^{-\alpha\phi}(dA^{a [D-4]}+\ast{\mathfrak G}^{a
[2]}+\alpha \ast (d\phi\wedge e^a)+e^{-\alpha\phi}\ast G^{a
[2]})\wedge (dz+A^{[1]}),
\end{align}
\begin{align}\label{Fz9s}
&\hat{\mathbb F}^{z [D-2]}=
a_{(1)}e^{-\beta\phi}(dA^{[D-3]}-{3\over 4} dA^{[D-4]}\wedge
A^{[1]})+a_{(1)}e^{2(D-3)\alpha\phi+\beta\phi} (1-\beta)\ast
d\phi\notag\\
&-a_{(1)}e^{2(D-3)\alpha\phi}\ast g^{z[1]}-a_{(2)} e^{-\beta
\phi}(dA^{[D-4]}+e^{-2\alpha\phi+\beta\phi}\ast G^{z[2]})\wedge
(dz+A^{[1]}).
\end{align}

Let us close this part of the paper with some remarks on the
proposed ansatz structure. First, the exponential factors of
leading expressions in \eqref{F8ans}, \eqref{F9ans},
\eqref{Fa8ans} are the same as in \eqref{KKe} but are opposite in
sign. This is the indication of duality. And second, there are
uncommon quantities entering the ansatz such as that of coming
from the reduction of $\hat{\tilde G}^{{\hat a}[2]}$. We are
forced to include such quantities into the ansatz since we can not
overcome the technical problem of evaluating the explicit
expression for $\hat{\tilde G}^{{\hat a}[2]}$ which is defined by
\eqref{tJ}. The same concerns to the ${\mathfrak G}^{a [2]}$.
However the ansatz we proposed does not depend on the explicit
structure of the above mentioned expressions and is universal,
though requires the very well knowledge of the duality relations
in lower dimensional theory after reduction.

\section{Duality-symmetric action for gravity and complete
duality-symmetric action for D=11 supergravity}

Having established the reduction ansatz let us turn to the more
sophisticated problem of constructing the action for the
duality-symmetric gravity. Since that requires more tuning we are
fixing the space-time dimensions $D$ to be eleven, and will follow
the notation of \cite{bns}. In this notation after dimensional
reduction we should in particular reproduce from the action the
duality relations \eqref{KKdr10}, \eqref{ddr10} which now have the
following form
\begin{equation}\label{F8dr}
{\mathcal F}^{[8]}=dA^{[7]}+e^{-{3\over 2}\phi}\ast dA^{[1]}=0,
\end{equation}
\begin{equation}\label{F9dr}
{\mathcal F}^{[9]}=(dA^{[8]}-{3\over 4}dA^{[7]}\wedge
A^{[1]})+\ast d\phi=0.
\end{equation}
Here we have fixed the parameters $\alpha$ and $\beta$ to be
\begin{equation}\label{abf}
\alpha=+{1\over 12},\qquad \beta=-{2\over 3}.
\end{equation}
The relevant part of the ten-dimensional duality-symmetric action
from which these relations come is (cf. \cite{bns})
\begin{equation}\label{dsL10}
S_{d.s.}={1\over 2} \int_{{\mathcal M}^{10}}\,
\sum_{n=1}^{2}\left(F^{[10-n]}\wedge F^{[n]}+i_v {\mathcal
F}^{[10-n]}\wedge v \wedge F^{[n]}+v\wedge F^{[10-n]}\wedge i_v
{\mathcal F}^{[n]}\right).
\end{equation}

Our aim is to demonstrate that the action \eqref{dsL10} follows
after dimensional reduction from the part of the gravity
duality-symmetric action
\begin{equation}\label{dsgL11}
S_{PST}=\int_{{\mathcal M}^{11}}\, {1\over 2} \hat{v}\wedge
\hat{\mathcal F}^{{\hat a}[9]}\wedge i_{\hat v} \hat{\mathcal
F}^{{\hat b}[2]}\eta_{{\hat a}{\hat b}},
\end{equation}
where following to the PST approach \cite{pst} we have introduced
a scalar field $a(\hat{x})$ that enters the action in a
non-polynomial way through the one-form ${\hat v}$
\begin{equation}\label{v11}
\hat{v}={d a(\hat{x})\over{\sqrt{-\partial_{\hat m}a~
\hat{g}^{{\hat m}{\hat n}}~\partial_{\hat n}a}}},
\end{equation}
$\eta_{{\hat a}{\hat b}}$ is the Minkowski metric tensor and
\begin{equation}\label{F2a11}
\hat{\mathcal F}^{{\hat a}[2]}=d\hat{e}^{{\hat
a}}-\hat{\ast}(d\hat{A}^{{\hat a}[8]}+\hat{\ast}\hat{\tilde
G}^{{\hat a}[2]}),
\end{equation}
\begin{equation}\label{F9a11}
\hat{\mathcal F}^{{\hat a}[9]}=-\hat{\ast}\hat{\mathcal F}^{{\hat
a}[2]}.
\end{equation}
Note that though these generalized field strengths are constructed
out of the non-covariant quantities, they enter the generalized
field strengths in the covariant combinations. Therefore,
${\mathcal F}^{a[2]}$ and ${\mathcal F}^{a[9]}$ are the covariant
objects.

Splitting the indices ${\hat a}$, ${\hat b}$ we can rewrite
\eqref{dsgL11} as
\begin{equation}\label{dsgL110}
S_{PST}=\int_{{\mathcal M}^{11}}\, {1\over 2} \hat{v}\wedge
\hat{\mathcal F}^{a [9]}\wedge i_{\hat v} \hat{\mathcal F}^{ b
[2]}\eta_{ab}-{1\over 2}\hat{v}\wedge \hat{\mathcal F}^{z
[9]}\wedge i_{\hat v} \hat{\mathcal F}^{ z [2]}.
\end{equation}
By use of ans\"atze \eqref{Fa9s}, \eqref{Fz9s} (in our notation
$a_{(1)}=a_{(2)}=-1$) one gets
\begin{equation}\label{Fa2red}
\hat{\mathcal F}^{a [2]}=e^{{1\over 12}\phi}[de^a+\ast (dA^{a
[7]}+\ast \mathfrak{G}^{a [2]})]\equiv e^{{1\over
12}\phi}{\mathcal F}^{a [2]},
\end{equation}
\begin{equation}\label{Fa9red}
\hat{\mathcal F}^{a [9]}=e^{-{1\over 12}\phi}{\mathcal F}^{a
[8]}\wedge (dz+A^{[1]}),\qquad {\mathcal F}^{a [8]}=\ast {\mathcal
F}^{a [2]},
\end{equation}
\begin{equation}\label{Fz9red}
\hat{\mathcal F}^{z [9]}=-e^{{2\over 3}\phi}({\mathcal
F}^{[9]}-{\mathcal F}^{[8]}\wedge (dz+A^{[1]})),
\end{equation}
\begin{equation}\label{Fz2red}
\hat{\mathcal F}^{z [2]}=e^{-{2\over 3}\phi}({\mathcal
F}^{[2]}-{\mathcal F}^{[1]}\wedge (dz+A^{[1]})),\quad {\mathcal
F}^{[9]}=\ast {\mathcal F}^{[1]},\quad {\mathcal
F}^{[8]}=e^{-{3\over 2}\phi}\ast {\mathcal F}^{[2]},
\end{equation}
where ${\mathcal F}^{[8]}$, ${\mathcal F}^{[9]}$ have been defined
in \eqref{F8dr}, \eqref{F9dr} (they are not equal to zero of
course; the latter shall follow from the action as an equation of
motion). Supplying the stuff by the following ansatz for $\hat{v}$
\cite{bns}
\begin{equation}\label{vans}
{\hat v}=e^{{1\over 12}\phi}v,\quad i_v (dz+A^{[1]})=0, \quad v={d
a(x)\over{\sqrt{-\partial_{m}a~ g^{mn}~\partial_{n}a}}},\quad v\ne
v(z),
\end{equation}
and using the standard rules of dimensional reduction (see e.g.
\cite{popelec} and Refs. therein, or Appendices in \cite{bns}),
after integration over $(dz+A^{[1]})$ one arrives at
\begin{equation}\label{Spst10}
S_{PST}=-{1\over 2} \int_{{\mathcal M}^{10}}\, v\wedge {\mathcal
F}^{a[8]}\wedge i_v {\mathcal F}^{b[2]}\eta_{ab}-v\wedge {\mathcal
F}^{[8]}\wedge i_v {\mathcal F}^{[2]}-v\wedge {\mathcal
F}^{[9]}\wedge i_v {\mathcal F}^{[1]}.
\end{equation}
After the standard manipulations one can convince yourself that
the reduction of the complete duality-symmetric D=11 gravity
action $S=S_{EH}+S_{PST}$ results in
\begin{align}\label{dsg10}
S=&-\int_{{\mathcal M}^{10}}\, \left(R^{ab}\wedge
\Sigma_{ab}+{1\over 2}v\wedge {\mathcal F}^{a[8]}\wedge i_v
{\mathcal
F}^{b[2]}\eta_{ab}\right)\notag\\
&+{1\over 2} \int_{{\mathcal M}^{10}}\,
\sum_{n=1}^{2}\left(F^{[10-n]}\wedge F^{[n]}+i_v {\mathcal
F}^{[10-n]}\wedge v \wedge F^{[n]}+v\wedge F^{[10-n]}\wedge i_v
{\mathcal F}^{[n]}\right).
\end{align}
It is easy to recognize the duality-symmetric structure of D=10
gravity action, and the second line is precisely the action
\eqref{dsL10} we are looking for. Therefore, we have established
the relevance of the proposed $S_{PST}$ term in the action for the
duality-symmetric gravity.

One may wonder why having a new term entering the action we did
not modify $\hat{\tilde G}^{{\hat a}[2]}$ though this term is not
a topological term and therefore contributes into the
energy-momentum tensor. The reason is the same as for another
application of the PST approach. The PST terms do not spoil the
dynamics of an original theory since their contribution to the
equations of motion is a combination of the duality relations
which is zero on-shell. The latter is guaranteed by a special
symmetry, one of the two PST symmetries \cite{pst}. That is for
instance why we do not need to take into account the contribution
of two last terms of \eqref{dsg10} into the ${\mathfrak
G}^{a[2]}$. Even leaving the issue of establishing the PST
symmetries out of consideration, one could notice an indication of
their presence in the eleven-dimensional theory since the special
symmetries of the last two terms of the action \eqref{dsg10} are
the bits of the corresponding eleven-dimensional symmetries.

To give more rigorous arguments in favor of the discussion above,
let us consider a variation of the action
\begin{equation}\label{AS}
S=S_{EH}+S_{PST}
\end{equation}
with $S_{EH}$ of \eqref{EH} and $S_{PST}$ of \eqref{dsgL11}. The
variation of the Einstein-Hilbert term results in
\begin{equation}\label{vEH}
\delta S_{EH}=\int_{{\mathcal M}^{11}}\, \delta {\hat e}^{\hat a}
\wedge d\hat{\mathcal F}^{{\hat b} [9]}\eta_{{\hat a}{\hat b}}.
\end{equation}
What concerns the PST term it is convenient to split its variation
into the standard for such an approach part where we will
effectively treat the vielbeins on the same footing as another
gauge field, i.e. as fields completely independent of metric, and
that of varying the metric
\begin{equation}\label{vPST}
\delta S_{PST}=\int_{{\mathcal M}^{11}}\, \delta_0 {\mathcal
L}_{PST}+\delta_{*}{\mathcal L}_{PST}.
\end{equation}

The standard variation ends up with
\begin{align}\label{v0PST}
\delta_0 S_{PST}&=\int_{{\mathcal M}^{11}}\, \left(\delta
\hat{A}^{{\hat a}[8]}+{\delta a \over{\sqrt{-(\partial a)^2}}}
i_{\hat v} \hat{\mathcal F}^{{\hat a}[8]}\right) \eta_{{\hat
a}{\hat b}} \wedge d(\hat{v}\wedge i_{\hat v}\hat{\mathcal
F}^{{\hat b}[2]})\notag\\
&+\int_{{\mathcal M}^{11}}\, \left(\delta \hat{e}^{\hat a}
+{\delta a \over{\sqrt{-(\partial a)^2}}} i_{\hat v} \hat{\mathcal
F}^{{\hat a}[2]}\right) \eta_{{\hat a}{\hat b}} \wedge
d(\hat{v}\wedge i_{\hat v}\hat{\mathcal
F}^{{\hat b}[9]})\notag\\
&-\int_{{\mathcal M}^{11}}\, \left(\delta \hat{e}^{\hat a}
\eta_{{\hat a}{\hat b}} \wedge d\hat{\mathcal F}^{{\hat
b}[9]}+\delta [\hat{\ast}\hat{\tilde{G}}^{{\hat a}[2]}]
\eta_{{\hat a}{\hat b}} \wedge \hat{v} \wedge i_{\hat v}
\hat{\mathcal F}^{{\hat b}[2]} \right).
\end{align}

To read off the other part of the variation we shall proceed as in
the case of extracting the energy-momentum tensor. In terms of
differential forms this variation is
\begin{align}\label{emn}
&\delta_* (\Omega^{[n]}\wedge \ast \Omega^{[n]})=(\delta_*
\Omega^{[n]})\wedge \ast \Omega^{[n]}+\Omega^{[n]}\wedge \delta_*
(\ast \Omega^{[n]})\notag\\
&={1\over (n-1)!}\delta e^{a_n}\wedge e^{a_{n-1}}\wedge\dots\wedge
e^{a_1} \Omega^{[n]}_{a_1 \dots a_{n-1} a_n}\wedge \ast
\Omega^{[n]}\notag\\
&+{1\over n!}{1\over (D-n-1)!}\Omega^{[n]}\wedge \delta
e^{b_{D-n}}\wedge e^{b_{D-n-1}}\wedge\dots\wedge
e^{b_1}{\epsilon_{b_1 \dots b_{D-n}}}^{a_1 \dots a_n}
\Omega^{[n]}_{a_1 \dots a_n}.
\end{align}
Following this pattern we arrive at
\begin{align}\label{v*PST}
\delta_* S_{PST}&=\int_{{\mathcal M}^{11}}\, {1\over 2} \hat{v}
\wedge \delta \hat{e}^{{\hat c}_9}({1\over 8!}\hat{e}^{{\hat
c}_8}\wedge\dots\wedge \hat{e}^{{\hat c}_1}{\hat{\mathcal
F}_{{\hat c}_1\dots {\hat c}_8 {\hat c}_9}}^{~{\hat
a}[9]})\eta_{{\hat a}{\hat b}}\wedge
i_{\hat v}\hat{\mathcal F}^{{\hat b}[2]}\notag\\
&+\int_{{\mathcal M}^{11}}\, {1\over 2} \hat{v}\wedge
\hat{\mathcal F}^{{\hat a}[9]}\eta_{{\hat a}{\hat b}} \wedge
\delta \hat{e}^{\hat c} (i_{\hat v} \hat{\mathcal F}^{{\hat
b}[2]})_{\hat c}.
\end{align}
Taking into account
%that $\hat{\ast} \hat{\tilde{G}}^{{\hat
%a}[2]}$ is a function of vielbeins, hence
%$$\delta [\hat{\ast} \hat{\tilde{G}}^{{\hat a}[2]}]=(\delta
%[\hat{\ast} \hat{\tilde{G}}^{{\hat a}[2]}]/ \delta \hat{e}^{\hat
%b})\wedge \delta \hat{e}^{\hat b},$$ and
the results of \eqref{vEH}, \eqref{v0PST}, \eqref{v*PST}, one can
derive the following sets of special symmetries
\begin{equation}\label{PST1}
\delta a(\hat{x})=0,\quad \delta \hat{e}^{\hat a}=da\wedge
\hat{\varphi}^{{\hat a}[0]},\quad \delta \hat{A}^{{\hat
a}[8]}=da\wedge \hat{\varphi}^{{\hat
a}[7]}-d^{-1}\delta(\hat{\ast}\hat{\tilde{G}}^{{\hat a}[2]}),
\end{equation}
\begin{align}\label{PST2}
&\delta a(\hat{x})=\Phi (\hat{x}),\qquad \delta \hat{e}^{\hat
a}=-{\Phi \over{\sqrt{-(\partial a)^2}}} i_{\hat v} \hat{\mathcal
F}^{{\hat a}[2]},\notag\\
&\delta \hat{A}^{{\hat a}[8]}=-{\Phi \over{\sqrt{-(\partial
a)^2}}} i_{\hat v} \hat{\mathcal F}^{{\hat
a}[9]}\notag\\&-d^{-1}\left(\delta(\hat{\ast}\hat{\tilde{G}}^{{\hat
a}[2]})+{1\over 2!8!}\delta\hat{e}^{{\hat c}_9}\wedge
\hat{e}^{{\hat c}_8}\wedge\dots\wedge \hat{e}^{{\hat
c}_1}\hat{\mathcal F}^{{\hat a}[9]}_{{\hat c}_1\dots {\hat
c}_8{\hat c}_9}-{1\over 2}\hat{\ast}(\delta \hat{e}^{\hat
c}\hat{\mathcal F}^{{\hat a}[2]})_{\hat c}\right).
\end{align}
Here we have introduced the inverse to $d$ operator whose action
on an arbitrary form is defined by use of a ``Green" function to
the equation
\begin{equation}\label{hdef}
d(x) h(x-y)=\delta^{11}(x-y),
\end{equation}
where $\delta^{11}(x)$ is the Dirac delta-function and therefore
\begin{equation}\label{d-1}
d^{-1}(x)\omega^{[p]}(x)=(-)^p \int \,
d^{11}y~h(x-y)\wedge\omega^{[p]}(y).
\end{equation}
To make a sense the latter expression should only deal with the
``Green" functions that act on a causality-related region of a
space-time.

It should be emphasized that we can not present the explicit
variation of the ``graviton dual" field without introducing the
non-local expressions since we are not having the explicit local
expression for $\hat{\ast} \hat{\tilde{G}}^{{\hat a}[2]}$. But
nevertheless, these non-local expressions do not spoil the job of
the PST symmetries \eqref{PST1}, \eqref{PST2} to extract the
duality relations and to establish the pure auxiliary nature of
the PST scalar field.

To demonstrate that it has to be noticed that the general solution
to the equation of motion of $\hat{A}^{{\hat a}[8]}$
\begin{equation}\label{A8eom}
d(\hat{v}\wedge i_{\hat v}\hat{\mathcal F}^{{\hat a}[2]})=0
\end{equation}
has the following form \cite{pst}
\begin{equation}\label{solA8eom}
\hat{v}\wedge i_{\hat v}\hat{\mathcal F}^{{\hat a}[2]}=da\wedge
d\hat{\xi}^{{\hat a}[0]}.
\end{equation}
Under the action of \eqref{PST1} the l.h.s. of the latter
expression is transformed as
\begin{equation}\label{lhstr}
\hat{v}\wedge i_{\hat v}\hat{\mathcal F}^{{\hat
a}[2]}\longrightarrow \hat{v}\wedge i_{\hat v}\hat{\mathcal
F}^{{\hat a}[2]}+da\wedge d\hat{\varphi}^{{\hat a}[0]},
\end{equation}
and therefore, setting $\hat{\xi}^{{\hat
a}[0]}=\hat{\varphi}^{{\hat a}[0]}$, one can reduce
\eqref{solA8eom} to
\begin{equation}\label{A8eomred}
i_{\hat v}\hat{\mathcal F}^{{\hat a}[2]}=0 \qquad\rightsquigarrow
\qquad \hat{\mathcal F}^{{\hat a}[2]}=0.
\end{equation}
Taking the latter into account and using the same arguments one
can reduce the vielbein equation of motion to
\begin{equation}\label{dr}
%\hat{\mathcal F}^{{\hat a}[2]}=0,\qquad
\hat{\mathcal F}^{{\hat a}[9]}=0.
\end{equation}
%from the equations of motion of $\hat{e}^{\hat a}$ and
%$\hat{A}^{{\hat a}[8]}$. Taking the latter into account
It becomes clear that the equation of motion of the PST scalar
$a(\hat{x})$ is identically satisfied as a consequence of the
equations of motion for the other fields. Therefore, the PST
scalar equation of motion is the Noether identity that is a
remnant of an additional symmetry which is nothing but the
symmetry under the transformations \eqref{PST2}.

Let us now extend the results to the bosonic sector of D=11
supergravity. To do that note first that as soon as the
three-index photon field $\hat{A}^{[3]}$ is taken into account eq.
\eqref{Fam} following from \eqref{Rem3} is replaced with
\begin{equation}\label{Famm}
\hat{F}^{[9]}_{\hat a}=d\hat{A}^{[8]}_{\hat
a}+\hat{\ast}\hat{\tilde{\mathbb G}}^{[2]}_{\hat a}\equiv
\hat{\mathbb F}^{[9]}_{\hat a}+\hat{\ast}\hat{\tilde{\mathbb
G}}^{[2]}_{\hat a},
\end{equation}
where $\hat{\tilde{\mathbb G}}^{[2]}_{\hat a}$ is defined by
$d\hat{\ast}\hat{\tilde{\mathbb G}}^{[2]}_{\hat
a}=\hat{\ast}\hat{\tilde{\mathbb J}}^{[1]}_{\hat a}$ with
$\hat{\ast}\hat{\tilde{\mathbb J}}^{[1]}_{\hat
a}=\hat{\ast}\hat{\tilde{J}}^{[1]}_{\hat a}+\hat{M}^{[10]}_{\hat
a}$. Here as in eq. \eqref{bfJd} we have introduced the
$\hat{A}^{[3]}$ energy-momentum form $\hat{M}^{[10]}_{\hat a}$.
After such a modification the reduction ans\"atze \eqref{Fa9s},
\eqref{Fz9s} shall be modified as follows
\begin{align}\label{Fa9s11}
&\hat{\mathbb F}^{a [9]}=e^{{4\over 3}\phi}\ast {\mathbf g}^{a
[1]}\notag \\
&+e^{-{1\over 12}\phi}(dA^{a [7]}+\ast\tilde{\mathfrak G}^{a
[2]}+{1\over 12} \ast (d\phi\wedge e^a)+e^{-{1\over 12}\phi}\ast
{\mathbb G}^{a [2]})\wedge (dz+A^{[1]}),
\end{align}
\begin{align}\label{Fz9s11}
&\hat{\mathbb F}^{z [9]}=-e^{{2\over 3}\phi}(dA^{[8]}-{3\over 4}
F^{[8]}\wedge A^{[1]}+{1\over 2}B^{[2]}\wedge dB^{[6]}-{1\over
4}F^{[6]}\wedge A^{[3]})\notag\\
&-e^{{2\over 3}\phi} (1+{2\over 3})\ast d\phi +e^{{4\over 3}
\phi}\ast {\mathbf g}^{z[1]}\notag\\
&+e^{{2\over 3}\phi}(dA^{[7]}+F^{[6]}\wedge B^{[2]}+B^{[2]}\wedge
B^{[2]}\wedge dA^{[3]}+e^{-{5\over 6}\phi}\ast {\mathbb
G}^{z[2]})\wedge (dz+A^{[1]}).
\end{align}
To present \eqref{Fa9s11}, \eqref{Fz9s11} we have used the
reduction rule $\hat{\tilde{\mathbb G}}^{[2]}_{\hat a}={\mathbb
G}^{[2]}_{\hat a}+{\mathbf g}^{[1]}_{\hat a}\wedge (dz+A^{[1]})$,
$\ast\tilde{\mathfrak G}^{a [2]}$ is the analog of the form
defined previously in \eqref{bfGd} and extended with the
appropriate contribution from the energy-momentum tensor of RR
3-form $A^{[3]}$ and of NS 2-form $B^{[2]}$. The latter forms come
from the reduction of $\hat{A}^{[3]}$. Finally, $F^{[6]}$ and
$F^{[8]}$ are the field strengths dual to the
$F^{[4]}=dA^{[3]}-dB^{[2]}\wedge A^{[1]}$ and $F^{[2]}=dA^{[1]}$
(see \cite{bns} for their explicit expressions).

Let us end up with the action for the complete duality-symmetric
bosonic sector of D=11 supergravity (cf. \cite{bbs}, \cite{bns}).
The action is as follows
\begin{align}\label{ds11sg}
S&=\int_{{\mathcal M}^{11}}\, \left[ \hat{R}^{{\hat a}{\hat
b}}\wedge \hat{\Sigma}_{{\hat a}{\hat b}}+{1\over
2}\hat{F}^{[4]}\wedge\hat{\ast}\hat{F}^{[4]}-{1\over
3}\hat{A}^{[3]}\wedge \hat{F}^{[4]} \wedge
\hat{F}^{[4]}\right]\notag\\
&+\int_{{\mathcal M}^{11}}\, {1\over 2}\left[ \hat{v}\wedge
\hat{\mathcal F}^{{\hat a}[9]}\wedge i_{\hat v} \hat{\mathcal
F}^{{\hat b}[2]}\eta_{{\hat a}{\hat b}}-\hat{v}\wedge
\hat{\mathcal F}^{[7]}\wedge i_{\hat v} \hat{\mathcal
F}^{[4]}\right],
\end{align}
with
\begin{equation}\label{F2a11s}
\hat{\mathcal F}^{{\hat a}[2]}=d\hat{e}^{{\hat
a}}-\hat{\ast}(d\hat{A}^{{\hat a}[8]}+\hat{\ast}\hat{\tilde
{\mathbb G}}^{{\hat a}[2]}),
\end{equation}
\begin{equation}\label{F9a11s}
\hat{\mathcal F}^{{\hat a}[9]}=-\hat{\ast}\hat{\mathcal F}^{{\hat
a}[2]},
\end{equation}
\begin{equation}\label{F4}
\hat{F}^{[4]}=d \hat{A}^{[3]},\qquad \hat{F}^{[7]}=d \hat
{A}^{[6]}+\hat{A}^{[3]}\wedge \hat{F}^{[4]},
\end{equation}
\begin{equation}\label{calF4}
\hat{\mathcal
F}^{[4]}=\hat{F}^{[4]}-\hat{\ast}\hat{F}^{[7]},\qquad
\hat{\mathcal F}^{[7]}=-\hat{\ast}\hat{\mathcal F}^{[4]}.
\end{equation}

One can verify that after dimensional reduction with taking into
account the ans\"atze \eqref{Fa9s11}, \eqref{Fz9s11} the action
\eqref{ds11sg} reproduces the following action of the completely
duality-symmetric type IIA supergravity (cf. eq. (69) in Ref.
\cite{bns})
\begin{align}\label{cdsg10}
&S=-\int_{{\mathcal M}^{10}}\, \left(R^{ab}\wedge
\Sigma_{ab}+{1\over 2}v\wedge {\mathcal F}^{a[8]}\wedge i_v
{\mathcal
F}^{b[2]}\eta_{ab}\right)\notag\\
&+{1\over 2} \int_{{\mathcal M}^{10}}\,
\sum_{n=1}^{4}\left({1\over 3^{[{n+1\over 4}]}} F^{[10-n]}\wedge
F^{[n]}+i_v {\mathcal F}^{[10-n]}\wedge v \wedge F^{[n]}+v\wedge
F^{[10-n]}\wedge i_v {\mathcal F}^{[n]}\right).
\end{align}

\section{Summary}

To summarize, we have exploited a similarity between the gravity
equation of motion and that of the Maxwell electrodynamics with an
electric-type source. The use of this fact allowed us to write
down the second order equation of motion for the vielbein as the
first order Bianchi identity for its dual partner. Knowing the
structure of the duality-symmetric action that follows from the
reduction of gravity action from $D$ to $D-1$ space-time
dimensions we have deduced the reduction ansatz for the field
strength of the ``graviton dual" field that reproduces the correct
$D-1$-dimensional duality relations between the dilaton and the
Kaluza-Klein vector field and their duals. After establishing the
ansatz we have proposed the additional PST-like term entering the
action of the duality-symmetric gravity and encoding the duality
relation between the graviton and its dual partner, and have
verified the relevance of this term by dimensional reduction from
eleven to ten comparing the results of reduction with that of
previously obtained in studying the duality-symmetric formulation
of type IIA supergravity. Finally, we have extended the ansatz to
the bosonic sector of D=11 supergravity and have established after
reduction the correct structure of the duality-symmetric type IIA
supergravity action. Therefore, we have non-trivially tested the
approach by Pasti, Sorokin and Tonin and have confirmed the
relevance of the approach to the construction of the
duality-symmetric supergravity actions in diverse dimensions.

In the course of our studying it has been also shown that the
duality relations for the graviton and its dual partner come as
equations of motion from the proposed action of the
duality-symmetric (super)gravity and that the PST scalar field
entering the action is an auxiliary field and does not spoil the
field content of an original theory. To demonstrate that we have
established the PST symmetries which are characteristic of the PST
approach and one of which shall be used to eliminate from a theory
the auxiliary PST scalar field. A feature of these transformations
consists in dealing with non-local terms. The same concerns the
action we proposed. In general there is not the exact local
expression for the Landau-Lifshitz ``pre-current" entering the
action, so the action is also constructed out of the non-local
terms. Having such terms is a feature of the self-interacting
theory since the double field approach to the Yang-Mills theory
also reveals non-locality \cite{ajnYM}.

Since we have not explicitly studied the fermionic sector here an
extension of the model to involve fermions remains conjectural at
this stage, but the experience in an immersion of PST formalism
into a supersymmetric theory lends credence to the conjecture.

\section*{Acknowledgments}
We are very grateful to Igor Bandos and Dmitri Sorokin for
valuable comments and encouragement. This work is supported in
part by the Grant \# F7/336-2001 of the Ukrainian SFFR and by the
INTAS Research Project \#2000-254.

\end{document}